\documentclass{ws-procs975x65}

\begin{document}

\title{MEASURING THE REDSHIFT OF STANDARD SIRENS USING THE NEUTRON STAR DEFORMABILITY}

\author{T. G. F. LI$^*$ and W. DEL POZZO}

\address{Nikhef - National Institute for Subatomic Physics,\\
Science Park 105, 1098 XG Amsterdam, The Netherlands\\
$^*$E-mail: tgfli@nikhef.nl}

\author{C. MESSENGER}

\address{School of Physics and Astronomy, Cardiff University,\\ 
Queens Buildings, The Parade, Cardiff, CF24 3AA, The United Kingdom}

\begin{abstract}
A recent study has shown that redshift information can be directly extracted
from gravitational wave sources. This can be done by exploiting the tidal
phasing contributions to the waveform during the inspiral phase of binary
neutron stars coalescences. The original study investigated the viability of
this idea in the context of the Einstein Telescope using a Fisher Matrix
approach and in this paper, we further explore this idea using realistic
simulations and Bayesian inference techniques. We find that the fractional
accuracy with which the redshift can be measured is in the order of tens of
percent, in agreement with Fisher Matrix predictions. Moreover, no significant
bias is found. We conclude that, when tidal phasing contributions are included
in the analysis, inference of the cosmological parameters from gravitational
waves is possible. 
\end{abstract}

\keywords{gravitational waves -- standard sirens -- neutron stars -- equation of state -- cosmology}

\bodymatter

\section{Introduction}
\label{sec:intro}
Gravitational waves (GWs) emitted by binary systems containing either neutron
stars (NSs) and/or black holes (BHs) directly encode information on the
distance between the source and the observer. A Hubble diagram, from which
cosmological information can be inferred, can be constructed if there is a
complementary redshift measurement for each source. However, GW signals from
such sources do not contain clear spectral lines as observed in electromagnetic
(EM) standard candles and will therefore rely on alternative methods for
redshift measurement. 

Messenger and Read have shown that redshift information can be directly
extracted from GWs by exploitation of the tidal phasing contributions to the
waveform in the inspiral phase of binary NS coalescences \cite{Messenger2012a}.
This additional phasing due to the tidal deformability of NSs breaks the
degeneracy between NS rest-mass and it's redshift, allowing both to be measured
simultaneously.  Moreover, they show that Einstein Telescope (ET) could measure
the redshift with an accuracy of order of tens of percent, depending on the NS
equation of state (EOS) that is assumed. These results were obtained through
Fisher matrix calculations and should be viewed as a lower bound on the
redshift accuracy when the signal-to-noise ratio (SNR) is high.

In this proceeding, we present the first realistic investigation of the
viability of this method using Bayesian inference techniques. In particular, we
show the sensitivity of ET to the redshift through a comprehensive simulation.

\section{Simulation}
\label{sec:simulation}

We restrict our sources to be binary NSs with component rest masses distributed
uniformly in the range $m\in \left[ 1,3 \right] M_{\odot}$, and with the EOS
knows as MS1\cite{Mueller1996a}. The binaries are distributed uniformly in sky
location, orientation, and co-moving volume with a maximum redshift of $z=4$.
We assume a concordance cosmology,
\textit{i.e.} $h = 0.7$, $\Omega_m = 0.3$, and $\Omega_\Lambda=0.7$. We
simulate GW signals using the frequency domain TaylorF2 waveform family for
non-spinning binaries \cite{Arun2005a}. The point particle description of the
phase in Ref.~\refcite{Arun2005a} is supplemented with the tidal contribution
given in Ref.~\refcite{Hinderer:2010kx}. Finally, we simulate the noise using
the one-sided power spectral density (PSD) labelled ET-B
\cite{2008arXiv0810.0604H}.

For each simulated GW event, we calculate the posterior probability density
function (posterior) for $z$, denoted by $p(z|d,I)$, where $d$ represents the
data and $I$ represents the information known prior to the arrival of the data.
The posterior for $z$ is calculated via marginalisation of the joint posterior over
the remaining GW parameters $\vec{\theta}$, \textit{i.e.} chirp mass, symmetric
mass ratio, right ascension, declination, polarisation, inclination, and time
and phase of coalescence:
\begin{align}
	p(z|d,I) = \int d\vec{\theta} \; p(z,\vec{\theta}|d,I) = \left[p(d|I)\right]^{-1} \int d\vec{\theta} \; p(d|z,\vec{\theta},I) p(z,\vec{\theta}|I).
	\label{eq:marginal_joint_posterior}
\end{align}
where the second equality follows from Bayes' theorem. Since we assume that the
cosmology is known, the luminosity distance is fixed by the redshift.

The likelihood of the data is given by
\begin{align}
	p(d|z,\vec{\theta},I) \propto \exp \left[ -2\Re \int_{f_0}^{f_{\rm LSO}} df \frac{\left| \tilde{d}(f) - \tilde{h}(z,\vec{\theta}; f) \right|^2}{S_n(f)}  \right],
	\label{eq:likelihood}
\end{align}
where $f_0=20\;\mathrm{Hz}$ is the lower cut-off frequency, $f_{\rm LSO}$ is
the frequency of the last stable orbit, $\tilde{d}(f)$ is the simulated data
(signal plus noise) in the frequency domain, $\tilde{h}(z,\vec{\theta}; f)$ is
template waveform for which we also use the TaylorF2 waveform family, and $S_n
(f)$ is the PSD. 

We assume that the prior distribution of the redshift is independent of all
remaining GW parameters allowing us to write $p(z,\vec{\theta}|I) = p(z|I)
p(\vec{\theta}|I)$. The prior on the GW parameters, $p(\vec{\theta}|I)$, is
taken to be a flat distribution over the range $m_i \in \left[ 1,3 \right]$
for the component rest masses, isotropic distributions for the sky location and
orientation the orbital plane, and flat distributions for the time and phase of
coalescences over the range $0\leq \phi_c \leq 2\pi$ and a time interval of
$100\,\textrm{msec}$ around the true time.  The prior on the redshift,
$p(z|I)$, is taken such that the sources are expected to be distributed uniformly
in co-moving volume, assuming the same cosmological parameters as for the source
simulation, up to a redshift of $z=4$. To perform the integral in
Eq.~\ref{eq:marginal_joint_posterior}, we use the implementation of the Nested
Sampling algorithm given in Ref.~\refcite{Veitch2010a}.

\section{Results \& Discussion}
\label{sec:results}
\begin{figure}[t]%
\begin{center}
\psfig{file=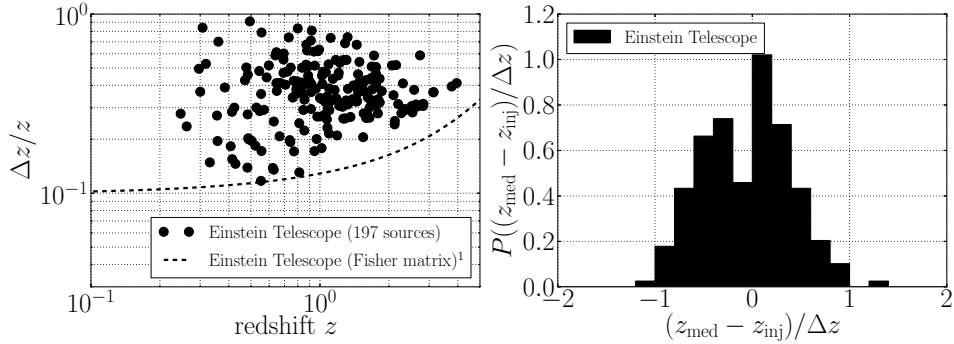,width=\textwidth}
 \caption{Left panel: Fractional error of the redshift as a function of the
 	 true redshift for the Einstein Telescope (circles), and the corresponding
 	 sky location and orientation-averaged Fisher matrix
 	 results\cite{Messenger2012a} (dashed line). The fractional error varies
 	 between 10-100 percent. Right panel: Distribution of fractional bias,
 	 $(z_{\rm med}-z_{\rm true}) / (\Delta z)$, where $z_{\rm med}$ is the median
 	 redshift and $z_{\rm true}$ is the true redshift. No systematic bias is
 found.}
\label{fig:redshift}
\end{center}
\end{figure}

The left panel of Fig.~\ref{fig:redshift} shows the fractional error of the
redshift, $\Delta z/z$ (where $\Delta z$ denotes the $68\%$ confidence
interval), for $197$ sources with a network SNR (ET comprises of three
co-located detectors) greater than 8. These results are compared to the Fisher
matrix calculations similar to those in Ref.~\refcite{Messenger2012a} but with
the ET-B PSD. In line with the Fisher Matrix calculations, the redshift can be
found with an accuracy of $\mathcal{O}(10^{-1})$, and the accuracy
decreases as the redshift increases.

The right panel of Fig.~\ref{fig:redshift} shows the distribution of the
fractional bias, $(z_{\rm med}-z_{\rm true}) / (\Delta z)$, where $z_{\rm med}$
is the median redshift and $z_{\rm true}$ is the true redshift, for the same
set of sources as the left panel. No systematic bias is found.

The results shown in Fig.~\ref{fig:redshift} suggest that it is indeed possible
to measure the redshift by supplementing the point-particle description of the
phase with corrections due to the NS tidal deformability. Whether the
accuracies shown in Fig.~\ref{fig:redshift} are sufficient to perform
competitive cosmological inference will be the subject of forthcoming
publications.

\bibliographystyle{ws-procs975x65}

\begin{thebibliography}{1}

\bibitem{Messenger2012a}
C.~Messenger and J.~Read, {\em Phys. Rev. Lett.} {\bf 108}, p. 091101 (Feb
  2012).

\bibitem{Mueller1996a}
H.~{M{\"u}ller} and B.~D. {Serot}, {\em Nuclear Physics A} {\bf 606}, 508 (Feb
  1996).

\bibitem{Arun2005a}
K.~G. {Arun}, B.~R. {Iyer}, B.~S. {Sathyaprakash} and P.~A. {Sundararajan},
  {\em Phys.~Rev.~D.} {\bf 71}, p. 084008(Apr 2005).

\bibitem{Hinderer:2010kx}
T.~{Hinderer}, B.~D. {Lackey}, R.~N. {Lang} and J.~S. {Read}, {\em
  Phys.~Rev.~D} {\bf 81}, p. 123016 (Jun 2010).

\bibitem{2008arXiv0810.0604H}
S.~{Hild}, S.~{Chelkowski} and A.~{Freise}, {\em ArXiv e-prints} (Oct 2008).

\bibitem{Veitch2010a}
J.~{Veitch} and A.~{Vecchio}, {\em Phys.~Rev.~D} {\bf 81}, p. 062003 (March
  2010).

\end{thebibliography}

\end{document}